\documentclass[12pt]{iopart}

\usepackage{epsfig,colordvi}
\usepackage{color}
\usepackage{here}

 \def\be{\begin{equation}}
 \def\ee{\end{equation}}
 \def\bea{\begin{eqnarray}}
 \def\eea{\end{eqnarray}}
 \def\bean{\begin{eqnarray*}}
 \def\eean{\end{eqnarray*}}
 \def\gsim{\mathrel{\rlap{\lower0.2em\hbox{$\sim$}}\raise0.2em\hbox{$>$}}}
 \def\ksim{\mathrel{\rlap{\lower0.2em\hbox{$\sim$}}\raise0.2em\hbox{$<$}}}
 \def\kg{\mathrel{\rlap{\lower0.25em\hbox{$>$}}\raise0.25em\hbox{$<$}}}

\begin{document}

\title{Tomography of the Quark Gluon Plasma by Heavy Quarks}

\author{P.B. Gossiaux  and J. Aichelin}

\address{SUBATECH, Universit\'e de Nantes, EMN, IN2P3/CNRS
\\ 4 rue Alfred Kastler, 44307 Nantes cedex 3, France}
\begin{abstract}
Using the recently published model \cite{Gossiaux:2008jv,goss2} for
the collisional energy loss of heavy quarks in a Quark Gluon Plasma
(QGP), based on perturbative QCD (pQCD), we study the centrality
dependence of $R_{AA}$ and $R_{AA}(p_T^{min})$,
measured by the Phenix collaboration, and compare our model with
other approaches based on pQCD and on Anti de Sitter/ Conformal
Field Theory (AdS/CFT).
\end{abstract}


\section{Introduction}
The analysis of the spectra of light hadrons, observed in
ultrarelativistic collisions of Au nuclei at a center of mass energy
of $\sqrt{s} = 200$ GeV, has revealed that in these collisions a new
kind of strongly interacting matter is produced. One of the
evidences is the observation that the spatial deformation of the
overlap zone of projectile and target, quantified by the
eccentricity $\epsilon$, is converted into an asymmetry in momentum
space in azimuthal direction, called elliptic flow
$v_2=<\cos{2(\phi-\phi_{reaction})}>$ \cite{ell}. The experimental
$v_2$ is quantitatively described by ideal hydrodynamics. This means
that the viscosity coefficient is small \cite{heko,teaney}. Even if
in the meantime a detailed analysis of the impact parameter
dependence of the elliptic flow of different particles has revealed
that the situation is a bit more complicated \cite{elstar} a
remarkable and unexpected degree of local thermalization is obtained
in this new kind of matter, the plasma of quarks and gluons (QGP).

This small viscosity has the unwanted consequence that a local
equilibrium among the constituents of the QGP, light quarks and
gluons, is maintained until the phase transition. Hence those
hadrons which contain only light quarks carry only information on
plasma properties close to the phase transition. Therefore, most of
the observed particles are not very useful to obtain the desired
information on the creation and time evolution of the QGP and one
has to concentrate on those few probes which do not come to an
equilibrium with the expanding QGP. These probes include photons,
jets and heavy mesons. The latter are an especially useful probe
because a) due to the large mass of the heavy quarks the kinematic
properties of heavy mesons are close to that of heavy quarks before
hadronization, b) the initial momentum distribution of heavy quarks
can be inferred from pp collisions and is therefore known.
Consequently, comparing the $p_T$ spectra of heavy mesons, obtained
in heavy ion reactions, with that of pp collisions one has direct
access to the momentum change which the heavy quarks suffer while
traversing the plasma because the cross section for collisions of
the heavy meson after hadronization is presumably small. For this
purpose one defines
$R_{AA}=d\sigma_{AA}/dp_T^2/(<N_c>d\sigma_{pp}/dp_T^2)$, where
$<N_c>$ is the average number of initial binary collisions. If heavy
quarks do not suffer from an energy loss while traversing the plasma
$R_{AA}$ should be $\approx1$ but this is not really true because
the transverse momentum of the partons which create the heavy quark
pair has been modified by the medium. This will be discussed below.

Because the mean free path of heavy quarks is shortest at the
beginning of the expansion, the deviation of $R_{AA}$ from one
encodes dominantly the interaction of the heavy quarks with the QGP
at the beginning of the expansion. Initially the heavy quarks are
isotropically distributed in azimuthal direction. They can get
elliptic flow only by interactions with the light quarks and gluons.
Because it takes time until the eccentricity is converted into
elliptic flow the $v_2$ of the heavy mesons is sensitive to the
interaction of the heavy quarks with the plasma at the end of the
expansion of the plasma.
\section{The Model}
Recently we have advanced a model \cite{Gossiaux:2008jv} which
studies the creation of heavy quarks in a QGP, their interaction
with the expanding plasma (described by ideal hydrodynamics) and how
this interaction modifies the observed spectra of heavy mesons (or
more precisely that of single non photonic electrons, the decay
products of heavy mesons). The elementary interaction between the
heavy quarks and the partons of the plasma, light quarks, q, and
gluons, g, is described by pQCD where the density, the temperature
and the average velocity of the partons is given by the
hydrodynamical expansion. The time evolution of the distribution of
the heavy quarks can either be calculated by a Boltzmann equation or
by a Fokker-Planck equation. The results presented here are based on
the solution of the Boltzmann equation and we use the Fokker Planck
approach only to calculate drag and diffusion coefficients which can
be compared with other approaches. The details of the model can be
found in ref.\cite{Gossiaux:2008jv,goss2}. As compared to former
approaches our approach differs in two respects:

\begin{itemize}
\item We employ a running nonperturbative coupling constant whose value
remains finite at $t \to 0$ \cite{Dokshitzer:1995qm}.
\item We use an infrared regulator in the t-channel which is
determined by hard thermal loop calculations, as done by Braaten and
Thoma \cite{Braaten:1991jj} in the case of QED. The details of how
to extend their approach to QCD is found in the appendix of ref.
\cite{Gossiaux:2008jv}.
\end{itemize}

\begin{figure}[H]
\begin{center}
\epsfig{file=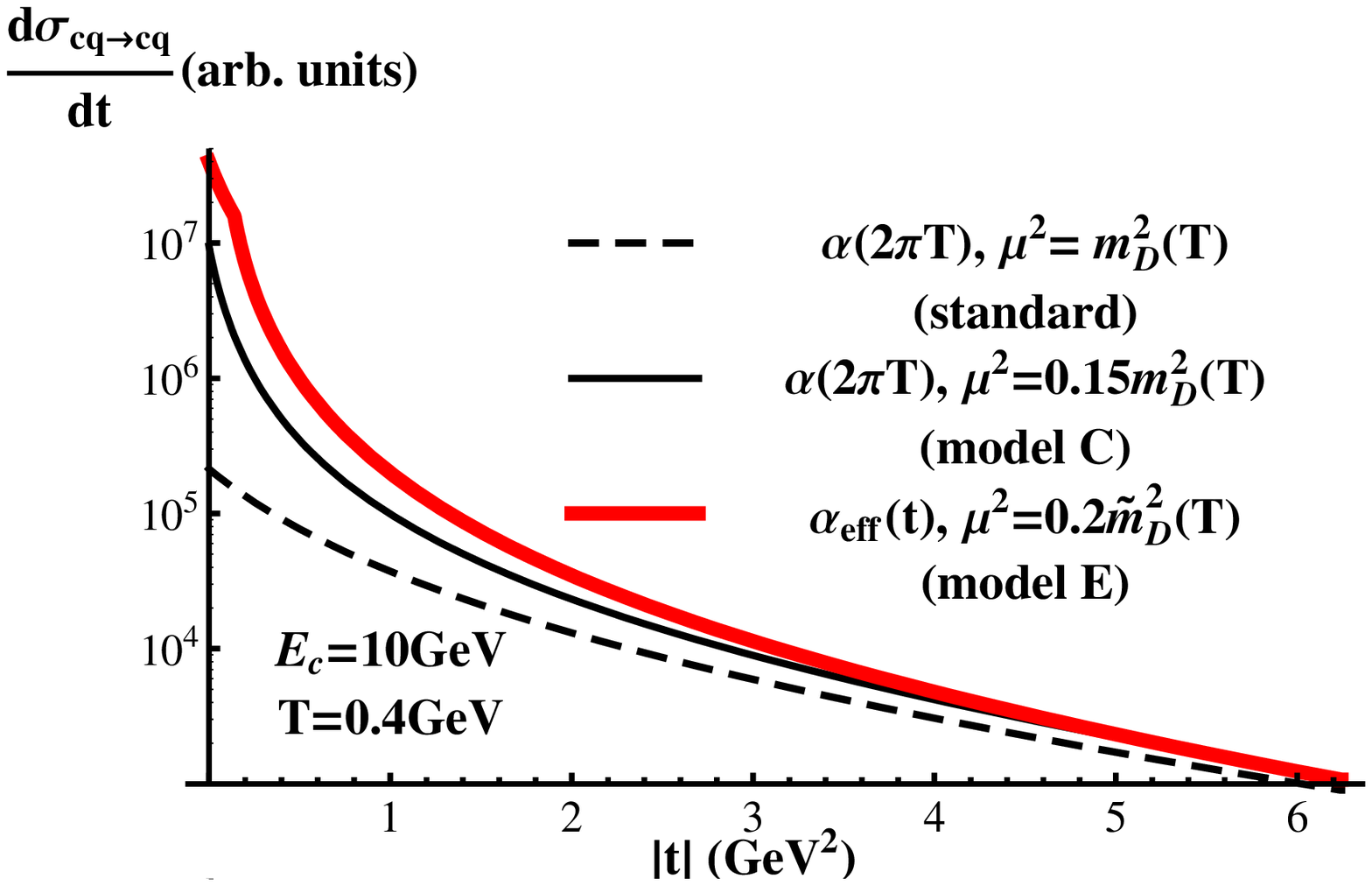,width=0.4\textwidth}
\epsfig{file=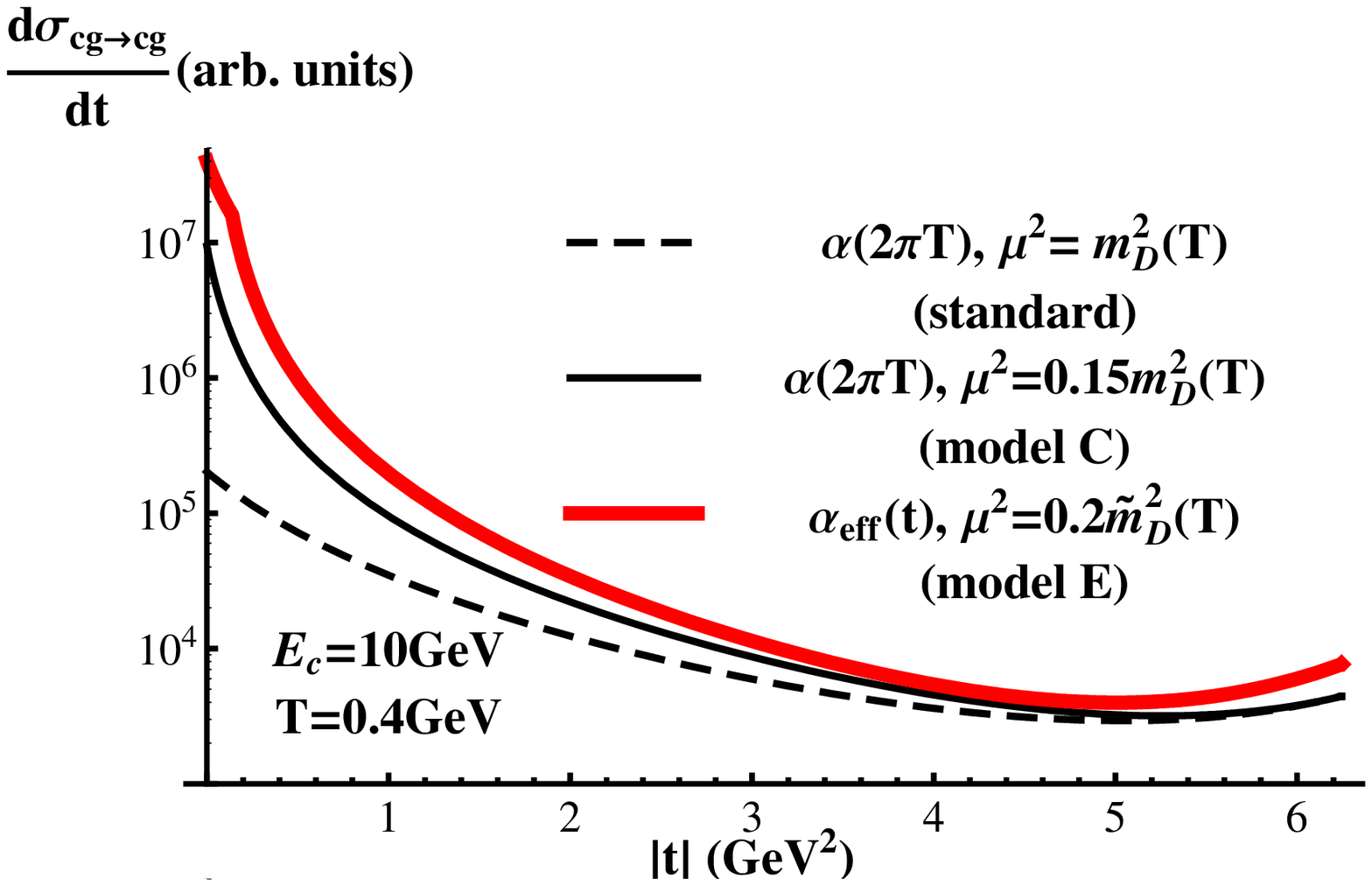,width=0.4\textwidth}
\end{center}
\caption{(Color online)Differential elastic cross section of $cq\to
cq$ (left) and $cg \to cg$ (right) for different choices of the
strong coupling constant and of the infrared regulator. The partons
are part of a heat bath of a temperature of 400 MeV and the c-quark
has an energy of 10 GeV.} \label{cro}
\end{figure}

Both these new ingredients enhance the elastic cross section in the
$qQ\to qQ$ as well as in the $gQ\to gQ$ channel. This can be seen in
fig. \ref{cro} which shows the total elastic cross section of a
c-quark with an energy of 10 GeV which traverses a  $T=400$ MeV
plasma, left for the collisions with light quarks, right for the
collisions with gluons for different assumptions on the coupling
constant and the infrared regulator. We see that the lower infrared
regulator as well as the running coupling constant increase the
cross section at low t as compared to the one with the standard
choices $\alpha(2\pi T)$ and $\mu=m_D$, where $m_D$ is the Debye
mass. Other approaches use a temperature independent coupling
constant and/or $\mu = k m_D$ where k varies between 0.3 and 1.
Energy loss by radiation is not taken into account yet in this
model. We introduce therefore a K-factor, i.e. a multiplication
factor which is applied to the cross section.  With a running
coupling constant and a infrared regulator of $\mu = 0.2m_D$, dubbed
"model E" in ref. \cite{Gossiaux:2008jv} and shown as the thick
(red) line in fig. \ref{cro}, a K-factor of 1.8 describes the
central as well as the minimum bias data for $R_{AA}$ and $v_2$
published by the STAR \cite{Abelev:2006db} and the Phenix
\cite{Adare:2006nq} collaboration, see ref.\cite{Gossiaux:2008jv}.

\begin{figure}[H]
\begin{center}
\epsfig{file=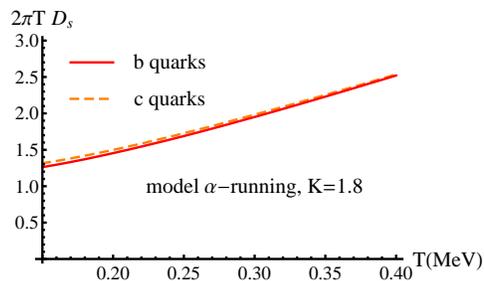,width=0.4\textwidth}
\end{center}
\caption{(Color online) $D_S$, the spatial diffusion coefficient for
c- and b-quarks, as a function of the temperature .} \label{spadif}
\end{figure}

In order to compare our model with other approaches we calculate the
diffusion constant in space, $D_S = <x^2(t)>/6t$, which is related
to the drag coefficient $\eta_D$ by $D_S=T/(M_Q \eta_D)$
\cite{Moore:2004tg}. The drag coefficient can be connected to the
ratio of viscosity and entropy density, $\eta$/s, one of the key
quantities of the present discussion. The relation is, however,
different in the different models and ranges from $\eta /s =D_ST/6$
\cite{Moore:2004tg} to $\eta /s =D_ST/2$ in the AdS/CFT approach.
 Fig. \ref{spadif} displays this quantity for b and c
-quarks as a function of the plasma temperature. $D_S$ is very
similar for c- and b-quarks in our approach and is for small
temperatures close to the quantal limit.
\section{Centrality dependence}
In ref.\cite{Gossiaux:2008jv} we have compared our results with
central and minimum bias data. As discussed in \cite{goss2} the
interaction between the heavy quarks and the plasma is a quite
complicated process in which the spatial geometry plays an essential
role and consequently the impact parameter (or centrality)
dependence of the results is highly non trivial. Therefore, it is
useful to exploit the whole selection of centralities provided by
the Phenix collaboration.

\begin{figure}
\begin{center}
\epsfig{file=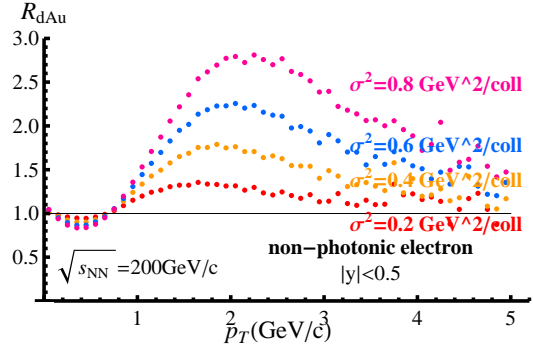,width=0.45\textwidth}
\end{center}
\caption{(Color online) $R_{dAu}$ as a function of $p_T$ for
different values of the variance $\sigma^2$ of the momentum
broadening due to the Cronin effect.} \label{dau}
\end{figure}

If hadrons have scattered before they create a heavy quark pair
their transverse momentum distribution is modified. This so called
Cronin effect yields a broadening of $\Delta P_T^2= n_{\rm
coll}(\vec{r}_{\perp})\,\sigma^2 $, where $\sigma^2$ is the
broadening of the squared transverse momentum in a single NN
collision and $n_{\rm coll}$ is the number of prior collisions. We
parameterize this distribution by a Gaussian function with a
variance of $\Delta P_T^2$. The consequences for $R_{dAu}$ for
different values of $\sigma^2$  are shown in fig. \ref{dau}. For
later calculations we use $\sigma^2 = 0.2\ {\rm GeV}^2$.

\begin{figure}
\begin{center}
\epsfig{file=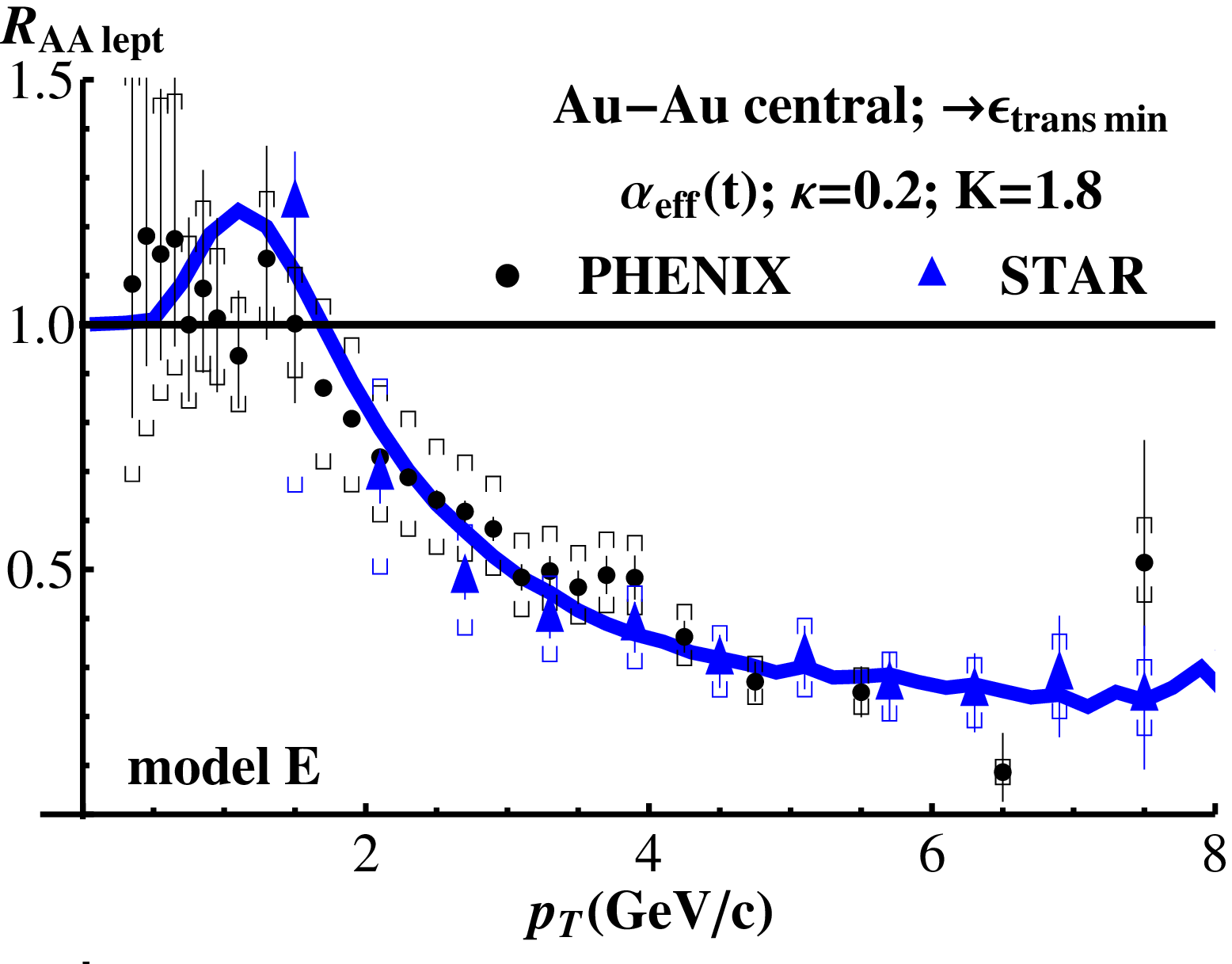,width=0.4\textwidth}
\epsfig{file=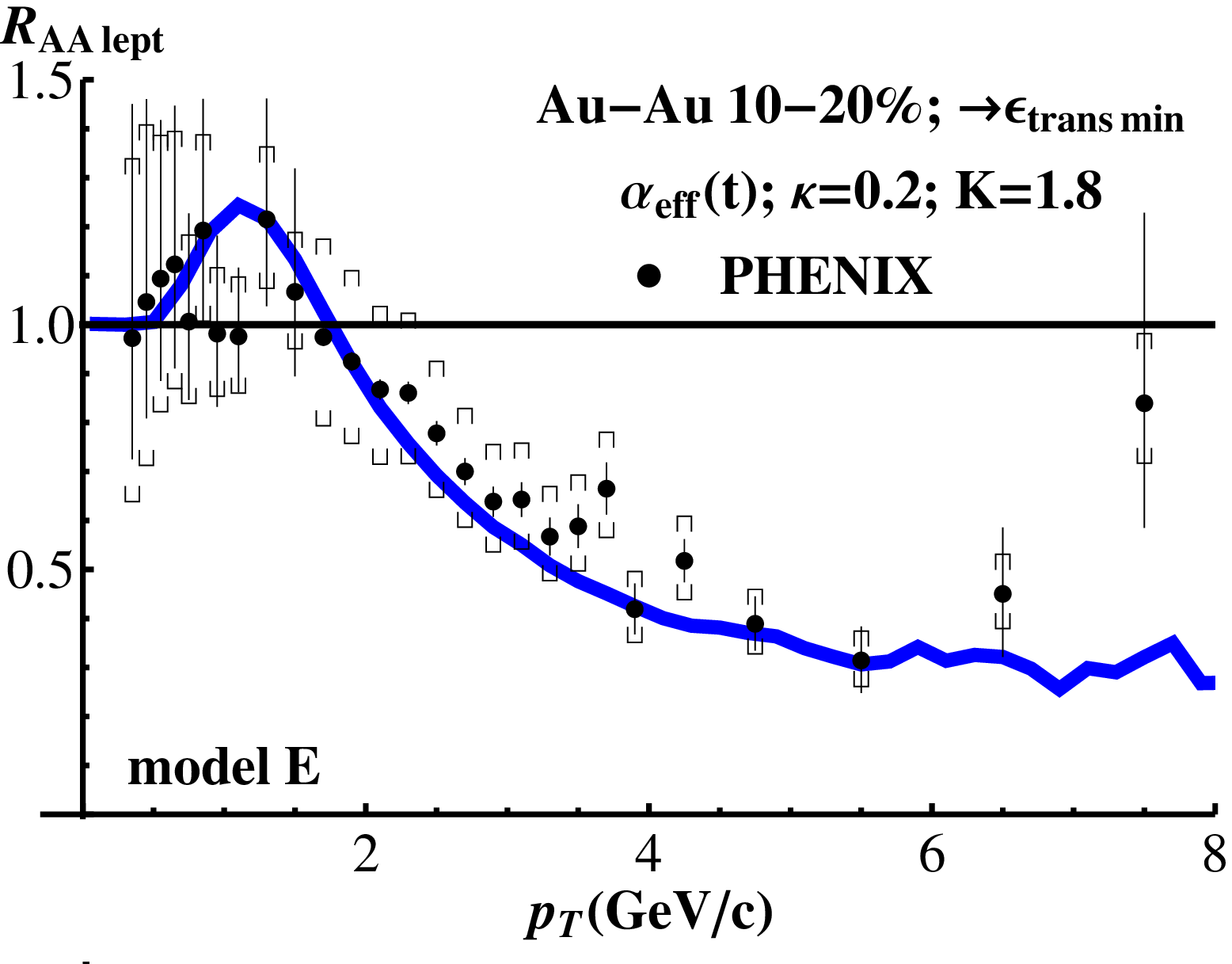,width=0.4\textwidth}
\epsfig{file=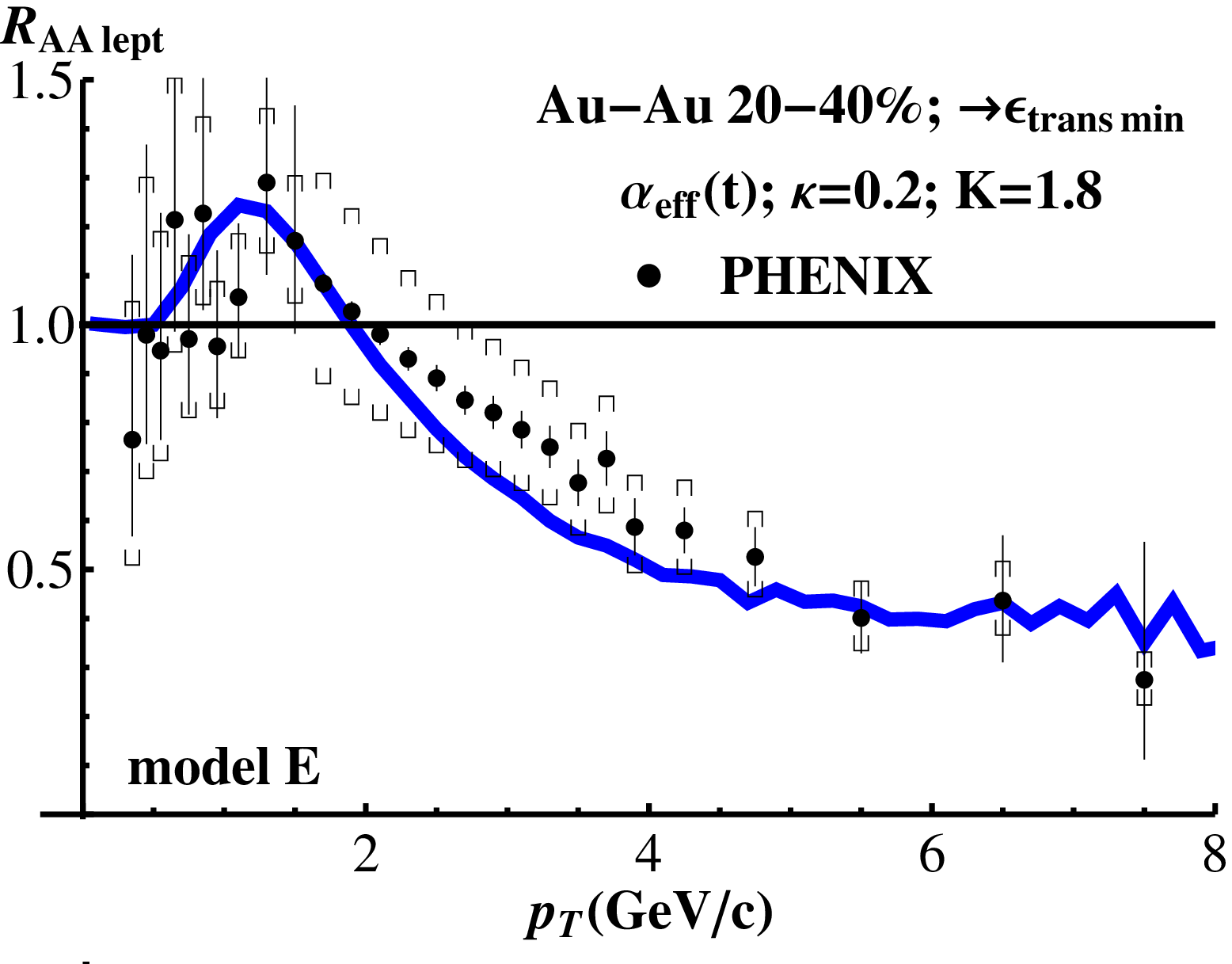,width=0.4\textwidth}
\epsfig{file=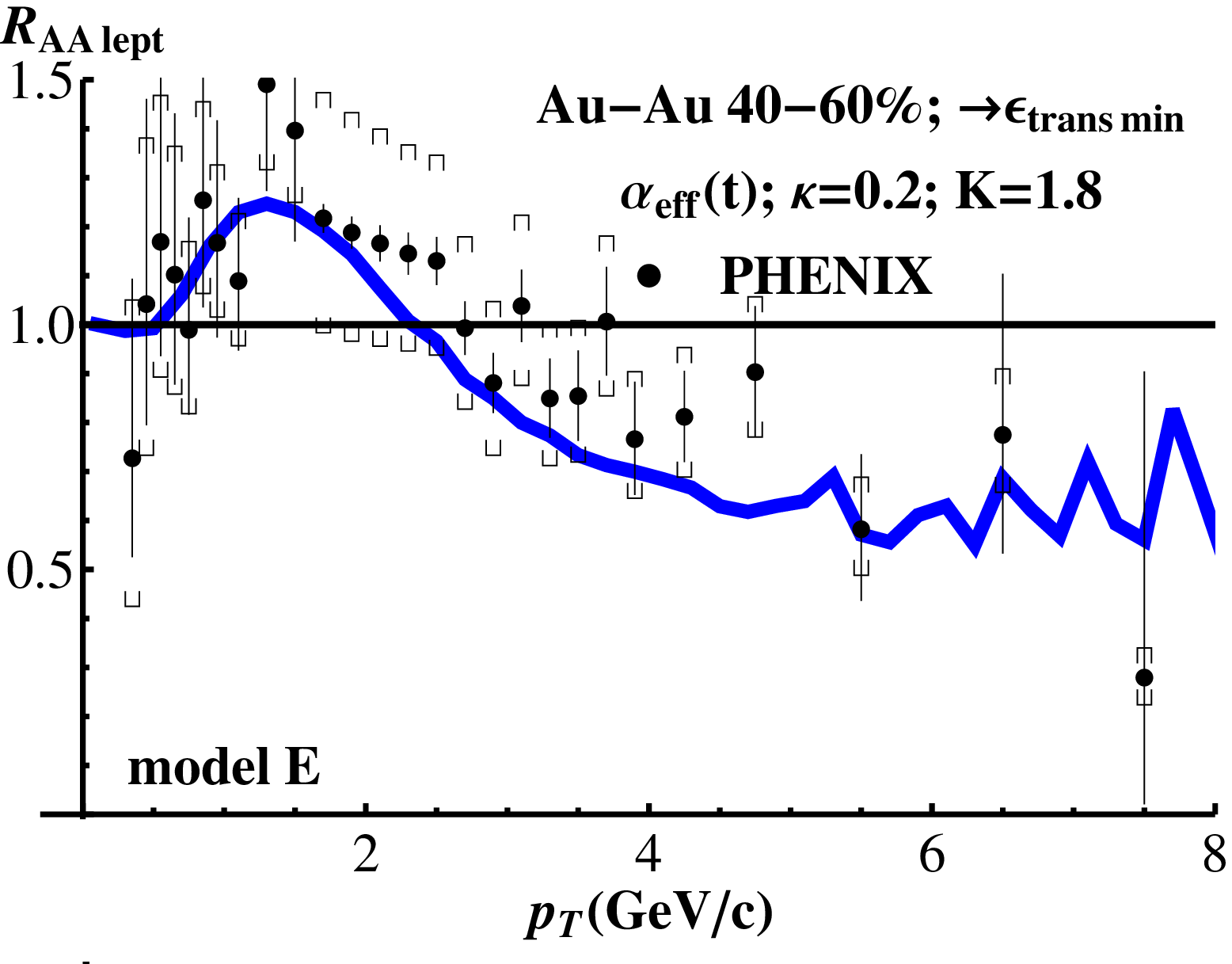,width=0.4\textwidth}
\epsfig{file=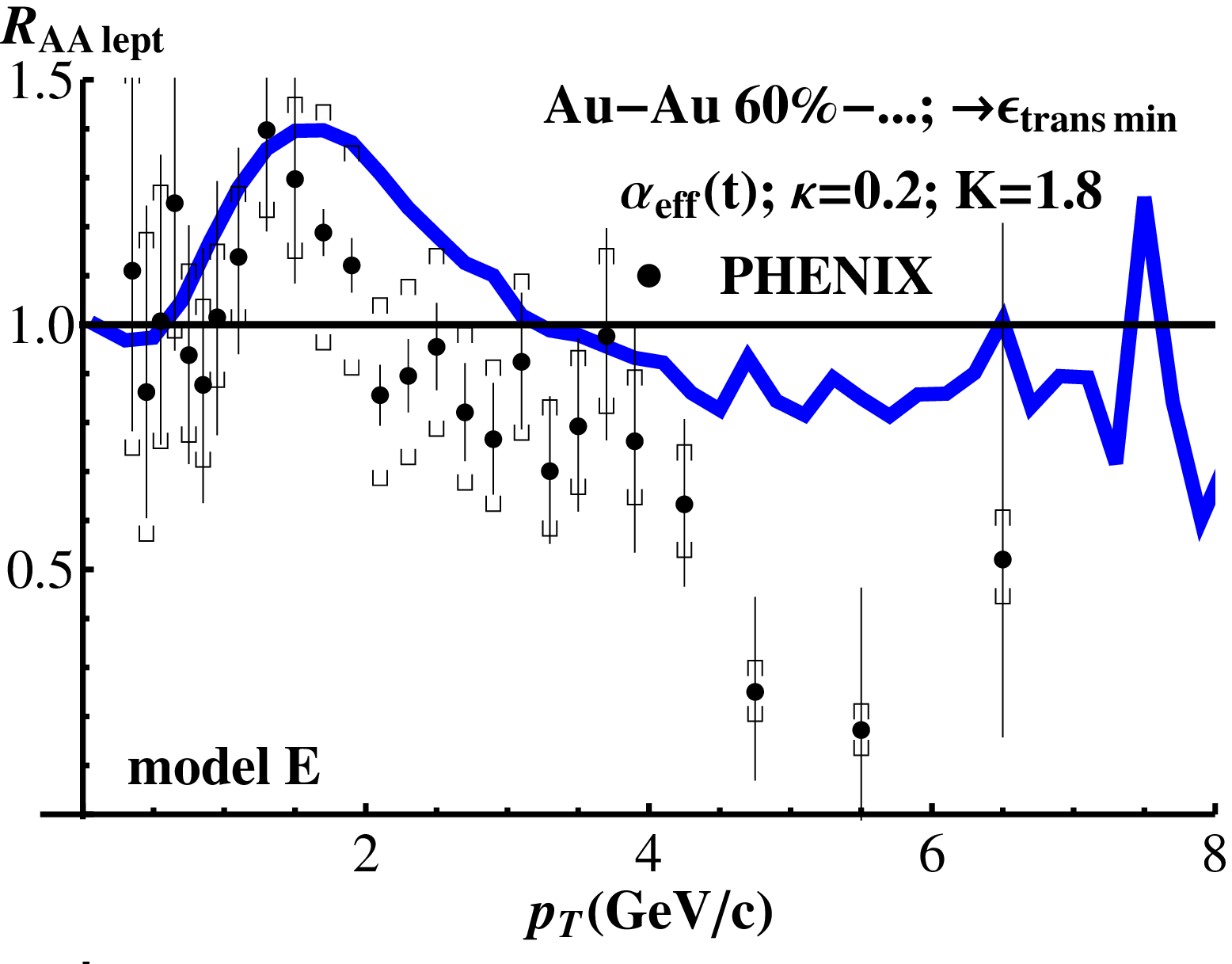,width=0.4\textwidth}
\epsfig{file=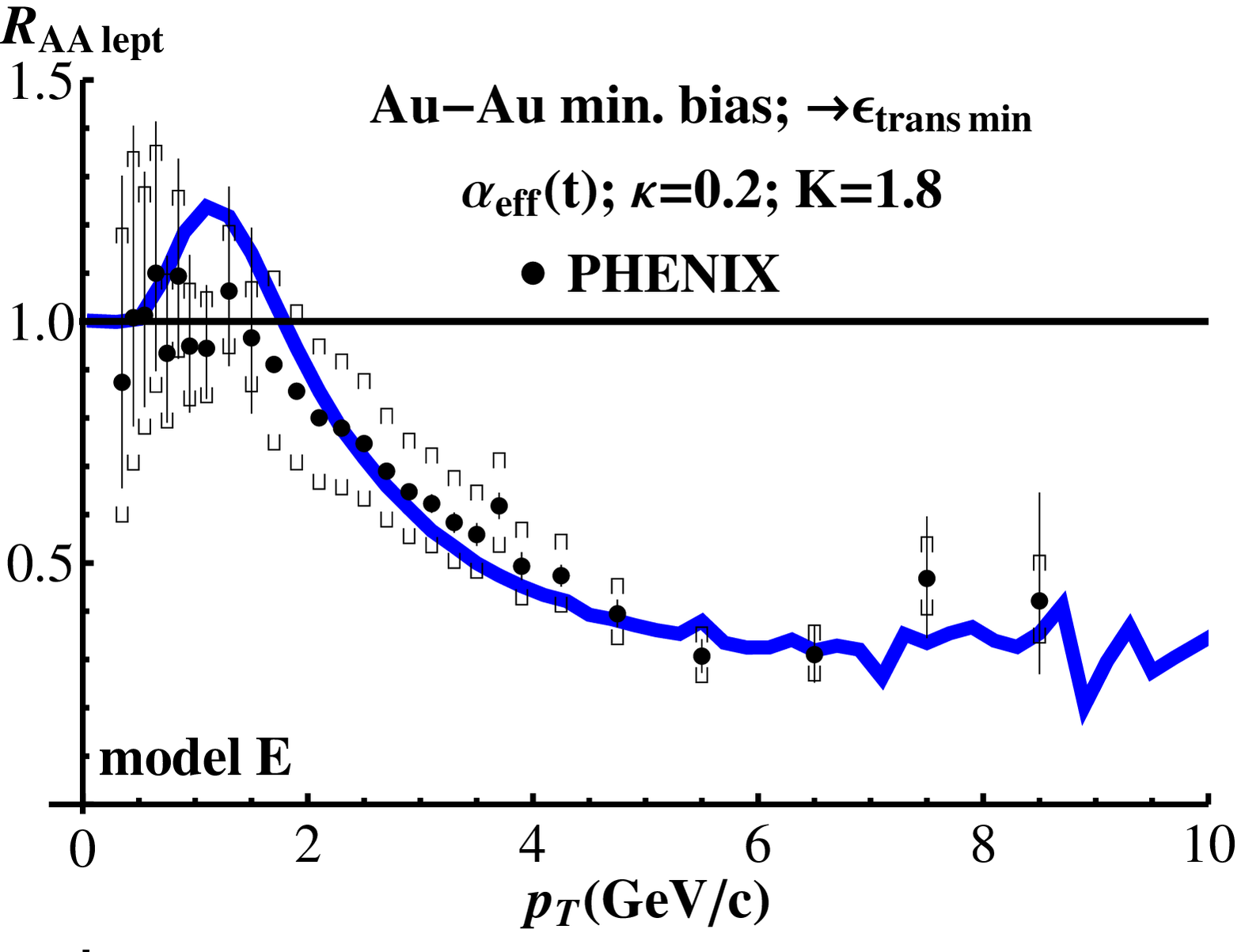,width=0.45\textwidth}
\end{center}
\caption{(Color online) $R_{AA}$ as a function of $p_T$ for
different centrality bins in comparison with PHENIX data
\cite{Adare:2006nq}. We display $R_{AA}$ for the model E
\cite{Gossiaux:2008jv} which uses a running coupling constant and an
infrared regulator determined by the hard thermal loop approach.}
\label{Phedata}
\end{figure}

The result of our approach for the different centrality classes, as
compared to the Phenix data, is shown in fig.\ref{Phedata}. We see
that for all centralities the general trend is well reproduced. For
the most peripheral events the Phenix data show an decrease of
$R_{AA}$ which is not reproduced in our approach and known
mechanisms do not account for this behavior.

\begin{figure}[H]
\begin{center}
\epsfig{file=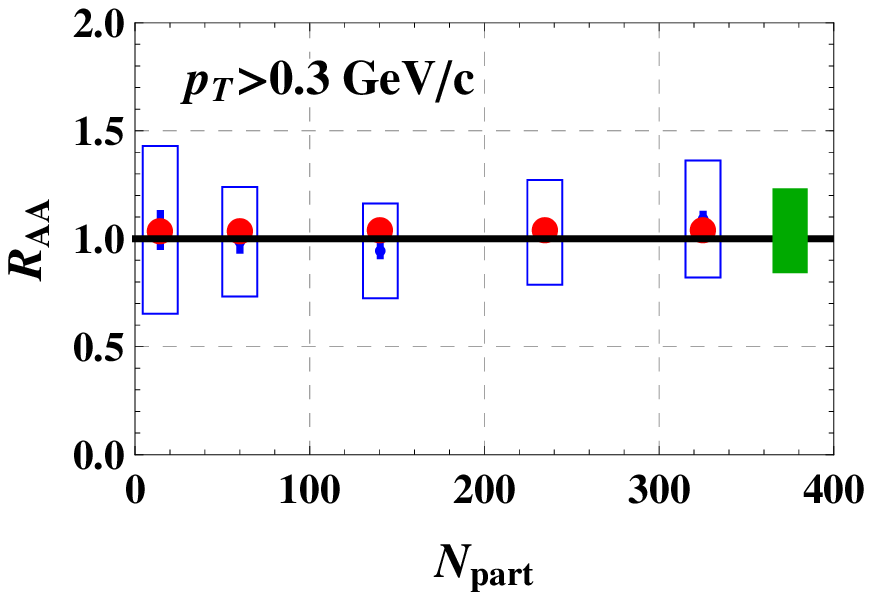,width=0.45\textwidth}
\epsfig{file=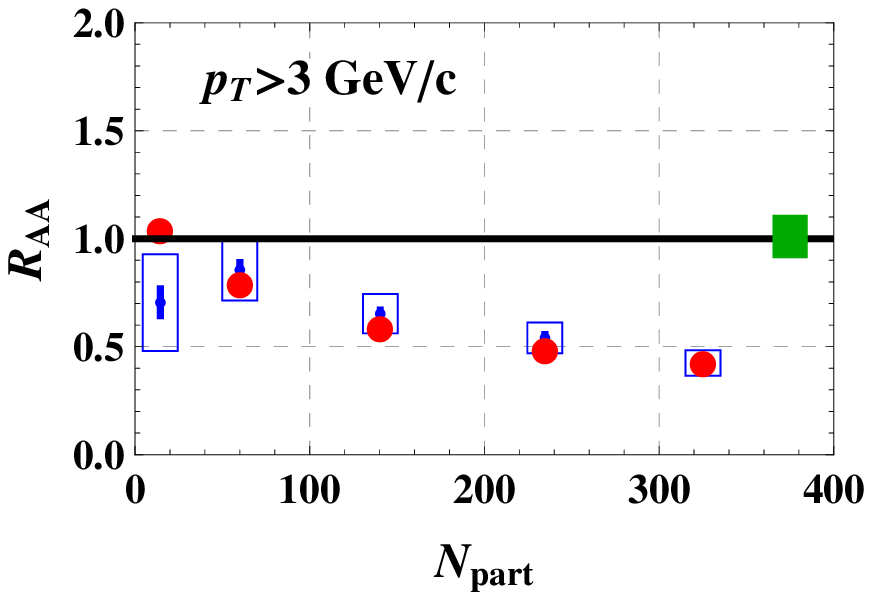,width=0.45\textwidth}
\end{center}
\caption{(Color online) $p_T^e$ integrated $R_{AA}$ (see text) as a
function of the centrality for different lower bounds of the
integration. We present the comparison of our model (dots) with the
data of the Phenix collaboration, presented as rectangles
\cite{Adare:2006nq}.} \label{Phecent}
\end{figure}

Another way to present the data is the centrality dependence of the
$p_T^e$ integrated $R_{AA}$ defined as \be
R_{AuAu}(p_T^{min})=\frac{\int_{p_T^{min}}^\infty
dN_{AuAu}/dp_T}{<N_c>\int_{p_T^{min}}^\infty dN_{pp}/dp_T}. \ee In
fig. \ref{Phecent} we present $R_{AuAu}(p_T^{min})$ as a function of
the participant number, $N_{part}$, and  for 2 different lower
bounds of the integration, $p_T^{min}$, in comparison with the
results of the Phenix collaboration \cite{Adare:2006nq}. Because all
heavy quarks are finally converted into heavy hadrons this
presentation allows to study  directly the average jet quenching as
a function of the centrality of the reaction.  Also these results
are in good agreement with the data.
\section{Comparison with other models}
\begin{figure}[H]
\begin{center}
\epsfig{file=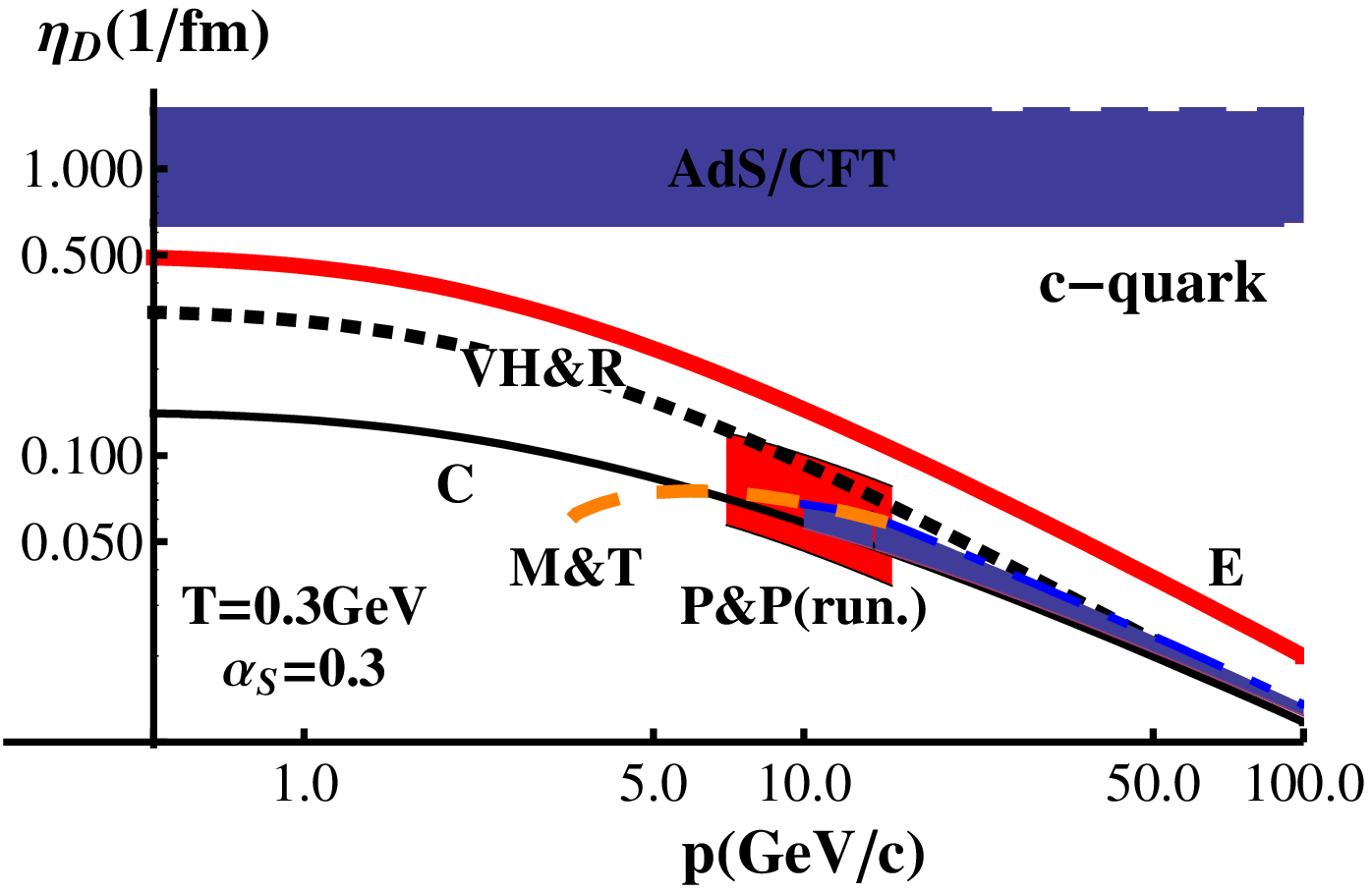,width=0.45\textwidth}
\epsfig{file=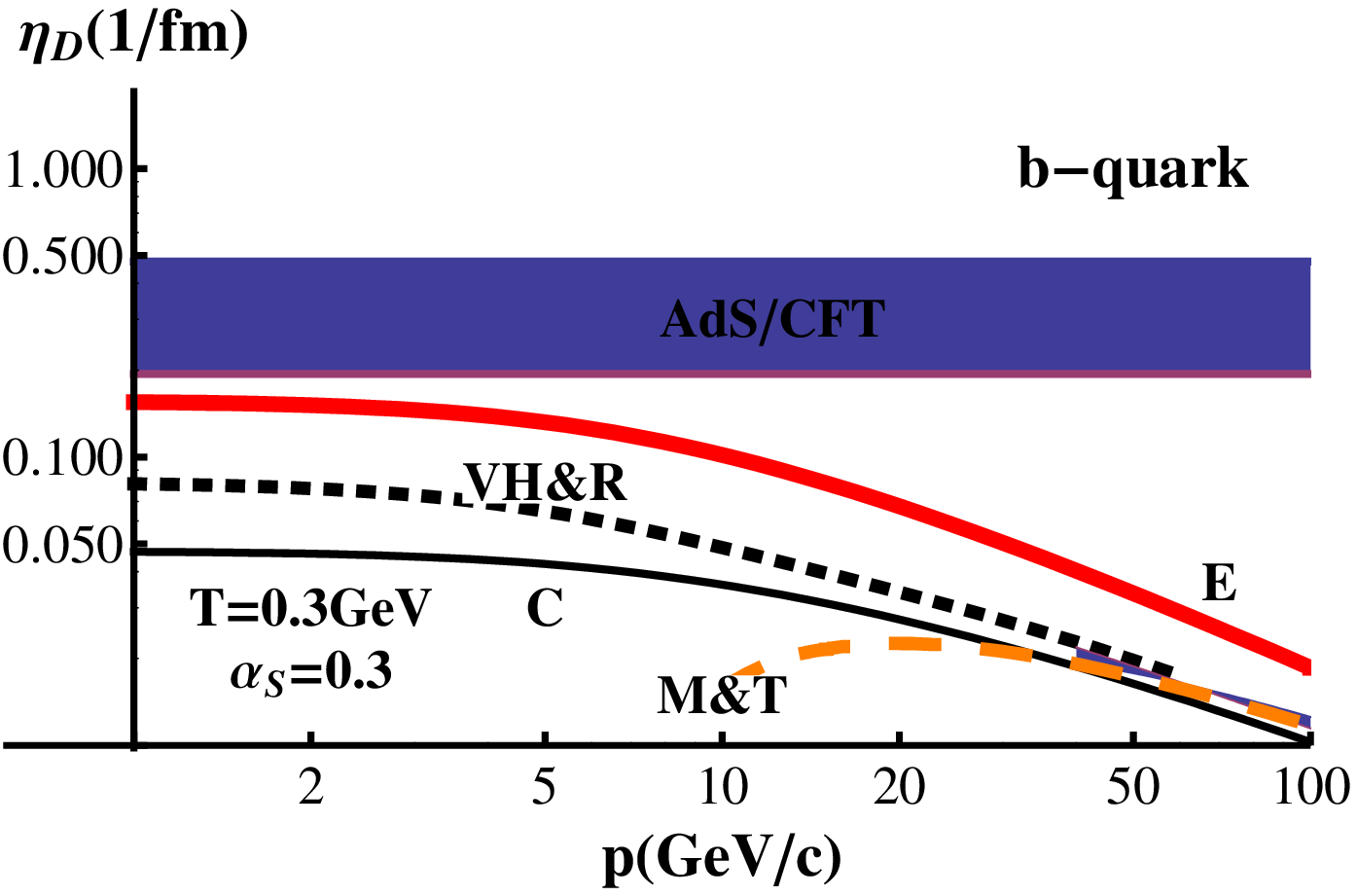,width=0.45\textwidth}
\end{center}
\caption{(Color online) The drag coefficient $\eta_D$ for c-quarks
(left) and b-quarks (right) as a function of the quark momentum. The
temperature of the scattering partners is 300 MeV.} \label{drag}
\end{figure}

\begin{figure}[H]
\begin{center}
\epsfig{file=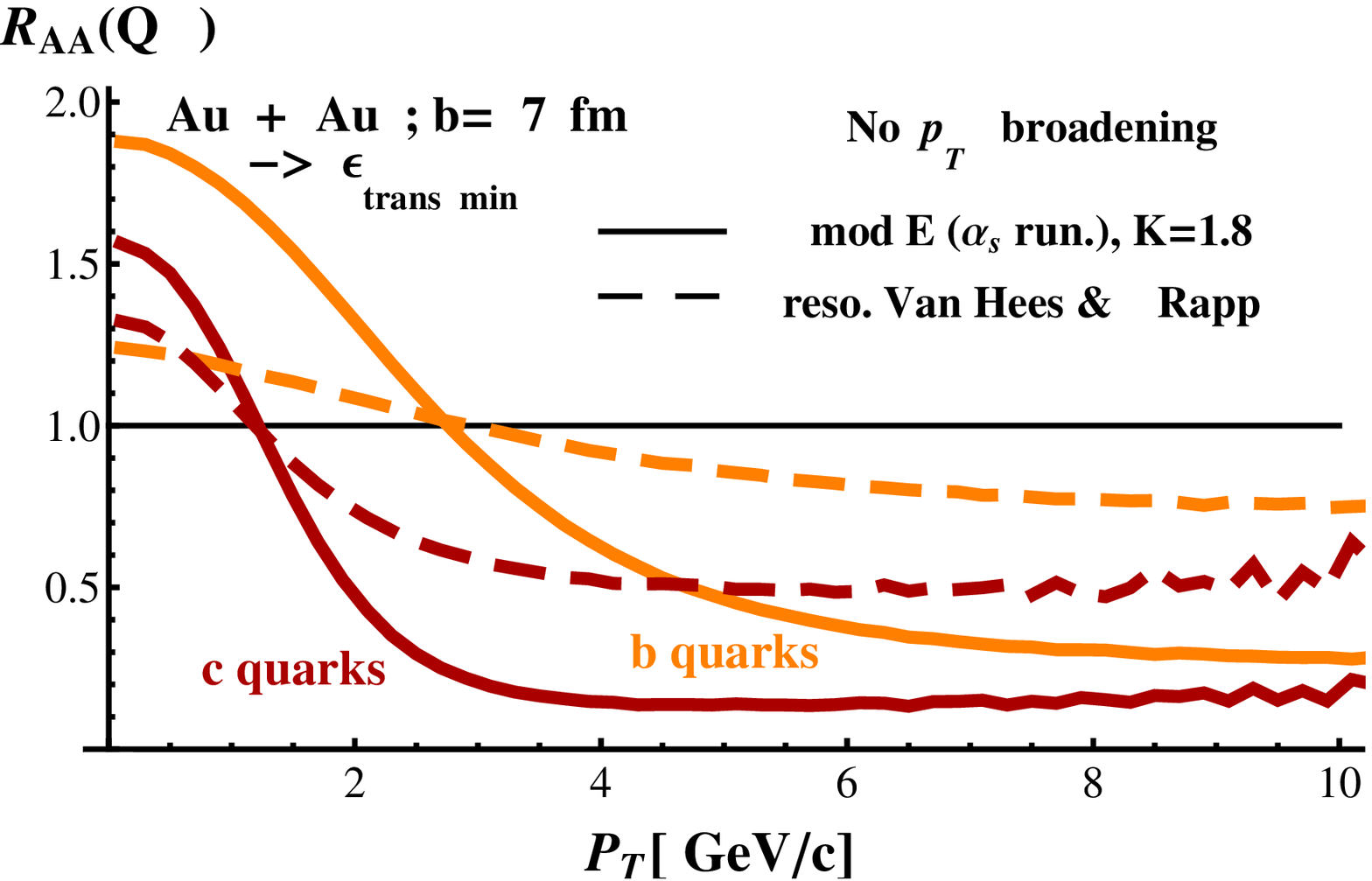,width=0.45\textwidth}
\epsfig{file=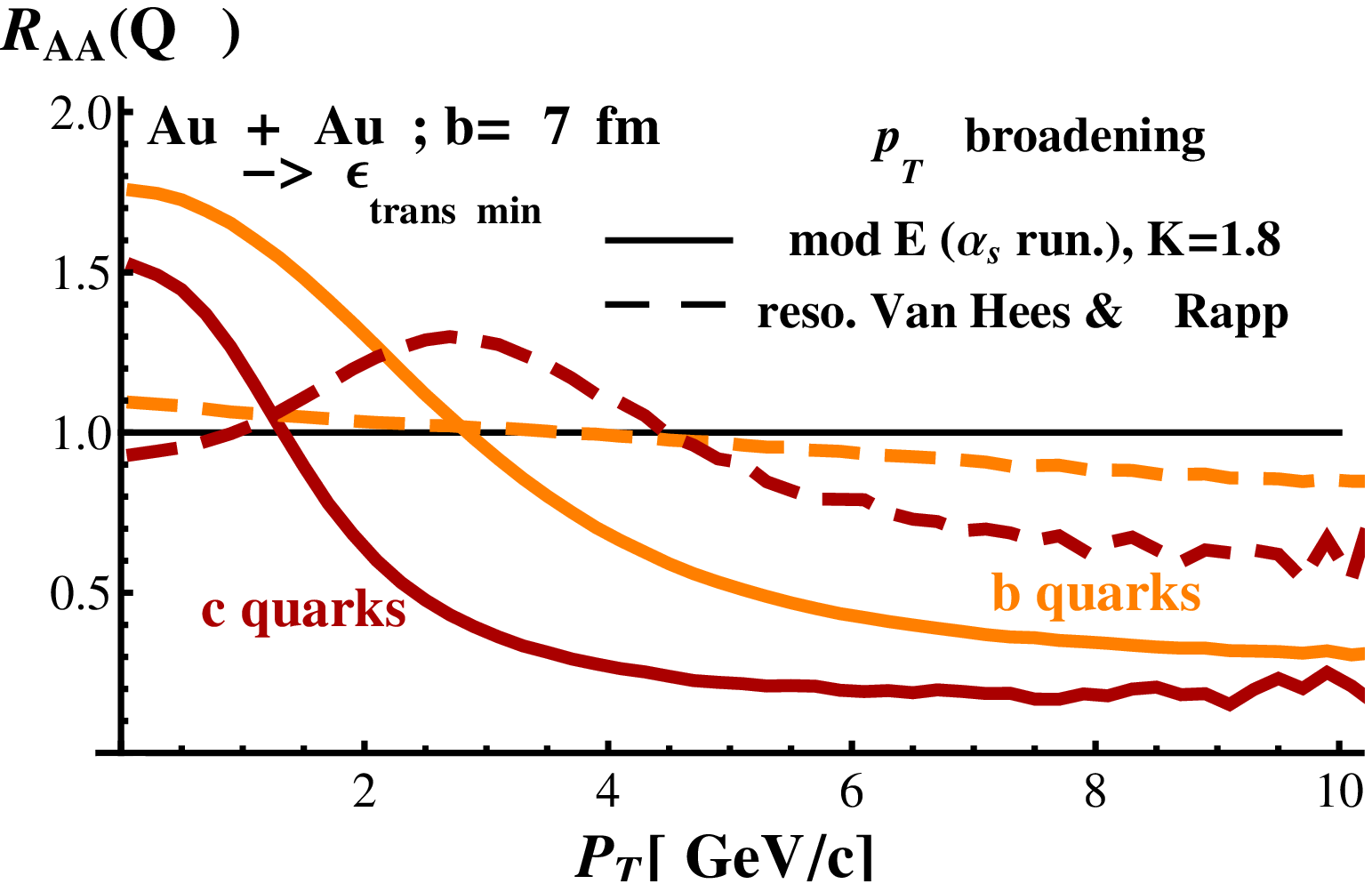,width=0.45\textwidth}
\epsfig{file=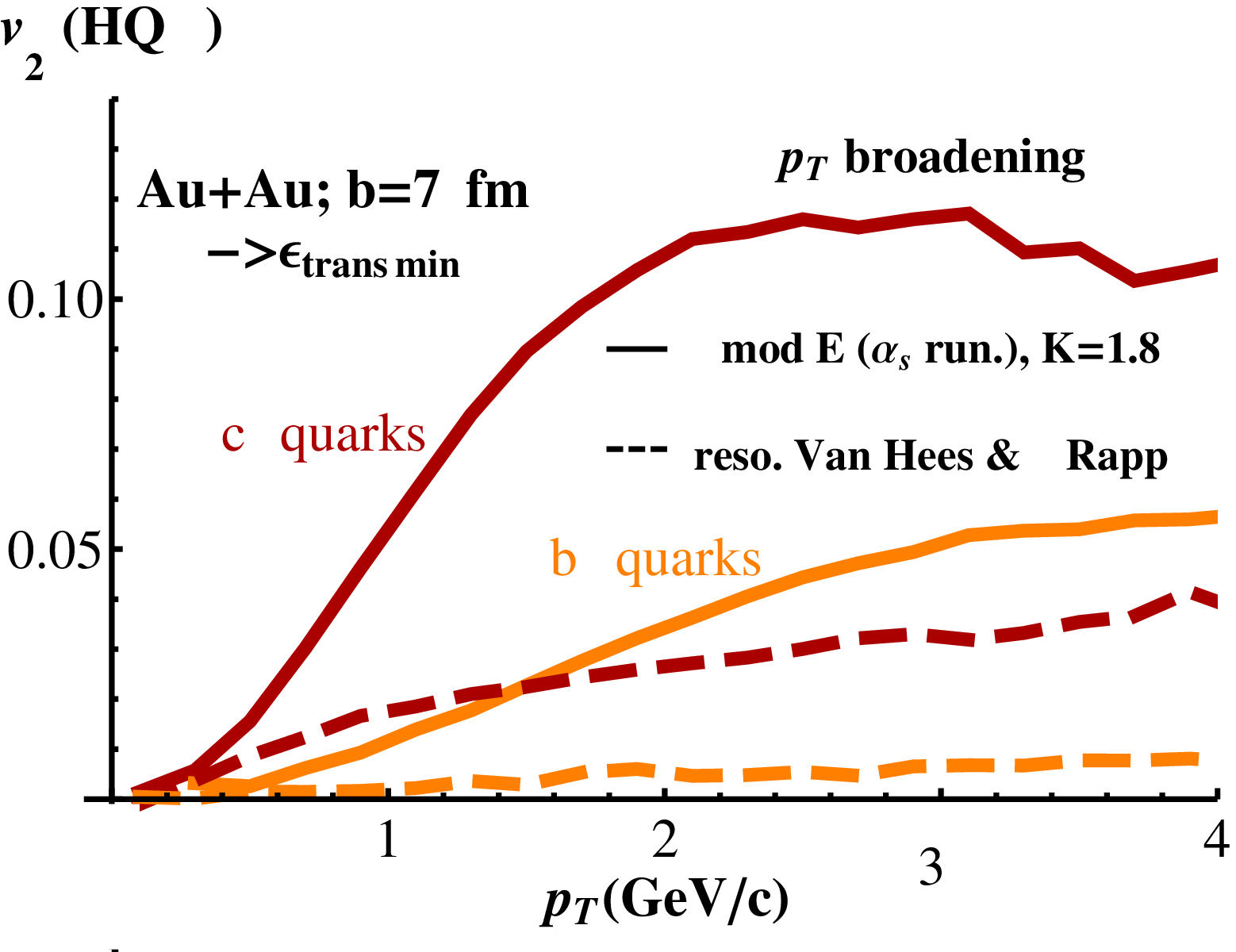,width=0.45\textwidth}
\end{center}
\caption{(Color online)$R_{AA}$ without (top,left) and with
(top,right) Cronin effect and $v_2$ (bottom) for Au+Au collisions at
$\sqrt{s}$=200 AGeV, b=7 fm as a function of $p_T$ for c- and b-
quarks and for two different approaches: our model E and a
calculation in which our drag coefficient has been replaced by that
of van Hees and Rapp.} \label{compvh}
\end{figure}

The theoretical values of the two experimentally measured
quantities, $R_{AA}$ and $v_2$, depend in a sensitive way on the two
conceptually different ingredients of the theory, the interaction of
the heavy quarks with the plasma and the expansion of the plasma
itself. If one wants to compare different theories it is very
helpful to separate both ingredients. This is possible by the
definition of transport coefficients which can be calculated in
every approach to the heavy quark-plasma interaction. These
transport coefficients depend on the interaction of the heavy quark
with the plasma but are independent of the expansion of the plasma.
The most interesting of these coefficients is the drag coefficient,
$\eta_D$. It describes the time evolution of the mean momentum
$\frac{dp}{dt}=\eta_D p$ of the heavy quark
\cite{reif,Cleymans:1985nd}. It can be calculated from the
microscopic interaction with help of eq. 2 of ref.
\cite{Gossiaux:2008jv} or from classical Langevin type approaches.
In some of the models it is given as an input variable. Fig.
\ref{drag} displays $\eta_D$ for different theoretical approaches,
on the left hand side for c-quarks and on the right hand side for
b-quarks. There we have assumed that the heavy quarks interact with
a plasma of a temperature of 300 MeV. For all calculations we use
the default values of the coupling constant. M\&T refers to Moore
and Teaney (eq. B31 with $\alpha_S = 0.3$) of \cite{Moore:2004tg},
VH\&R to van Hees and Rapp \cite{vanHees:2004gq,van
Hees:2005wb,Greco:2007sz} (with a resonance width of $\Gamma = 400$
MeV), P\&P to Peshier and Peigne \cite{Peigne:2008nd} and AdS/CFT to
the drag coefficient calculated in the framework of the anti de
Sitter/ Conformal field theory by Gubser
\cite{Horowitz:2008ig,Gubser:2006qh}. C (with $\alpha_S = 0.3$) and
E refer to two parameter sets of our model, defined in
\cite{Gossiaux:2008jv}.
The largest drag we observe for the AdS/CFT approach. In this theory
the drag coefficient is momentum independent. All (p)QCD based drag
coefficients decrease with increasing $p_T$ and hence with
increasing momentum the plasma becomes more transparent.
Nevertheless, the pQCD based drag coefficient vary quite
substantially due to different assumptions on the cut-offs and due
to different ingredients, especially the presence of qQ resonances
in the plasma. One may ask the question whether such difference does
not have a consequences on the predictions of experimentally
accessible quantities. In order to study this question we compare
the results of two calculations: those of our model E with those in
which the drag coefficient of model E is replaced by that of van
Hees and Rapp. The expansion of the plasma, described by the
hydrodynamical approach of Kolb and Heinz \cite{heko}, is identical
in both calculations. Fig. \ref{compvh} shows the results. The top
panels display $R_{AA}$ as a function of $p_T$ for c- and b-quarks,
left (right) without (with) $p_T$ broadening due to the Cronin
effect. In our model, which agrees with the data, the deviation of
$R_{AA}$ from one is twice as large for large $p_T$ as if we use the
drag coefficient of van Hees and Rapp in an otherwise unchanged
model. A similar observation can be made for $v_2$, see bottom
panel. For the impact parameter which has been used to simulate
minimum bias events, the elliptic flow $v_2$ is reduced by a factor
of two if we replace in our model the drag coefficient by that of
van Hees and Rapp without changing the model for the plasma
expansion. In their original publication van Hees and Rapp have
described the data quite well. It is therefore interesting to
explore whether the different models for the expansion of the plasma
are at the origin of the difference. If this were the case it would
stress another time the fact that the description of the
experimental $R_{AA}$ and $v_2$ spectra is a double challenge: that
to describe the Qq and Qg interactions and that to describe the
expansion of the plasma. If this were not the case, the heavy quarks
would not tell us something about the plasma properties {\it during}
the expansion.

\ack We would like to acknowledge valuable discussions with van Hees
and Rapp and to thank them for making their drift and diffusion
coefficient available to us.


\begin{thebibliography}{11}
\bibitem{Gossiaux:2008jv}
  P.~B.~Gossiaux, J.~Aichelin
  Phys.\ Rev.\   {\bf C78}, 014904 (2008), arXiv:0802.2525
  [hep-ph]
\bibitem{goss2}P.~B.~Gossiaux, R.~ Bierkandt and J.~Aichelin
  arXiv:0901.0946 [hep-ph]
\bibitem{ell}
  Ollitrault J-Y 1998
  {\it Nucl. Phys. } {\bf A638} 195c \\
  Poskanzer A M and Voloshin S A 1998
  {\it Phys. Rev.} {\bf C58} 1671
\bibitem{elstar}
B.~I.~Abelev {\it et al.}  [STAR Collaboration], arXiv:0801.3466
\bibitem{heko}
P. Kolb and U. Heinz, in Quark Gluon Plasma, World Scientific
Singapore, ed R.~Hwa and X.N.~Wang
\bibitem{teaney}
D. Teaney
  {\it Phys. Rev.} {\bf C68} 034913 (2003)
\bibitem{Peshier:2006ah}
  A.~Peshier,
  arXiv:hep-ph/0601119
\bibitem{Dokshitzer:1995qm}
  {}Y.~L.~Dokshitzer, G.~Marchesini and B.~R.~Webber
  {}Nucl.\ Phys.\  B {\bf 469}, 93 (1996)
  [arXiv:hep-ph/9512336]
\bibitem{Braaten:1991jj}
  {}E.~Braaten and M.~H.~Thoma
  {}Phys.\ Rev.\  D {\bf 44}, 1298 (1991),
  E.~Braaten and M.~H.~Thoma,
  Phys.\ Rev.\ D {\bf 44} (1991) 2625.
\bibitem{Abelev:2006db}
  B.~I.~Abelev {\it et al.}  [STAR Collaboration],
  Phys.\ Rev.\ Lett.\  {\bf 98}, 192301 (2007). 
\bibitem{Adare:2006nq}
  A.~Adare {\it et al.}  [PHENIX Collaboration],
  Phys.\ Rev.\ Lett.\  {\bf 98}, 172301 (2007)  [arXiv:nucl-ex/0611018].
\bibitem{reif} F. Reif "Fundamentals of Statistical Physics",
  McGraw-Hill (1965)
\bibitem{Cleymans:1985nd}
  J.~Cleymans and P.~S.~Ray
  BI-TP 85/08(1985)
  \bibitem{Moore:2004tg}
  G.~D.~Moore and D.~Teaney,
  Phys.\ Rev.\  C {\bf 71} 064904 (2005).

\bibitem{vanHees:2004gq}
  H.~van Hees and R.~Rapp,
  Phys.\ Rev.\  C {\bf 71}, 034907 (2005).
  [arXiv:nucl-th/0412015].


\bibitem{van Hees:2005wb}
  {}H.~van Hees, V.~Greco and R.~Rapp
  {}Phys.\ Rev.\  C {\bf 73}, 034913 (2006)
  [arXiv:nucl-th/0508055]

\bibitem{Greco:2007sz}
  {}V.~Greco, H.~van Hees and R.~Rapp
  {}arXiv:0709.4452 [hep-ph]
\bibitem{Peigne:2008nd}
  {}S.~Peigne and A.~Peshier
  {}Phys.\ Rev.\  D {\bf 77}, 114017 (2008)
  [arXiv:0802.4364 [hep-ph]]
\bibitem{Horowitz:2008ig}
  {}W.~A.~Horowitz and M.~Gyulassy
  {}arXiv:0804.4330 [hep-ph]

\bibitem{Gubser:2006qh}
  {}S.~S.~Gubser
  {}Phys.\ Rev.\  D {\bf 76}, 126003 (2007)
  [arXiv:hep-th/0611272]
\end{thebibliography}
\end{document}